\documentclass[twocolumn,showpacs,preprintnumbers,amsmath,amssymb]{revtex4}

\usepackage{graphicx}
\usepackage{dcolumn}
\usepackage{bm}

\begin{document}

\preprint{APS/123-QED}

\title{Local spin and charge properties of $\beta$-Ag$_{0.33}$V$_2$O$_5$ studied by $^{51}$V-NMR }

\author{A. Hisada and N. Fujiwara}
 \email{ naokif@mbox.kudpc.kyoto-u.ac.jp}

\affiliation{%
 Graduate School of Human and Environmental Studies, Kyoto
 University, Yoshida-nihonmatsu-cho, Sakyo-ku, Kyoto 606-8501, Japan
}%

\author{T. Yamauchi and Y. Ueda}

\affiliation{%
 Institute for Solid State Physics, University of Tokyo, 5-1-5
 Kashiwanoha, Kashiwa, Chiba, 277-8581, Japan
}%


\date{\today}

\begin{abstract}
 Local spin and charge properties were studied on $\beta$-Ag$_{0.33}$V$_2$O$_5$, a pressure-induced superconductor,
 at ambient pressure using $^{51}$V-NMR and zero-field-resonance (ZFR) techniques. Three inequivalent Vi
 sites (i=1, 2, and 3) were identified from $^{51}$V-NMR spectra and
 the principal axes of the electric-field-gradient (EFG) tensor were
 determined in a metallic phase and the following charge-ordering phase. We found from the EFG analysis that the V1 sites are in
 a similar local environment to the V3 sites. This was also observed
 in ZFR spectra as pairs of signals closely located with each other.
 These results are well explained by a charge-sharing model where a 3d$^1$ electron is shared within a rung in both V1-V3 and
 V2-V2 two-leg ladders.
\end{abstract}

\pacs{76.60.-k; 74.70.-b}
\maketitle

\section{INTRODUCTION}
 A family of $\beta$-vanadium bronze, $\beta$-A$_{0.33}$V$_2$O$_5$ (A=Li,Na,Ag), has offered an attractive stage
 in strongly correlated electron systems because it exhibits various phases
 on pressure-temperature (\emph{P-T}) phase diagram; a metallic phase at high temperatures,
 an charge-ordering (CO) phase followed by an antiferromagnetic (AF) phase
 at low temperatures,$^{1-4}$ and pressure-induced superconductivity at pressures above 7GPa.$^{5-8}$
 The superconducting (SC) phase is adjacent to the CO phase at lower pressure region.
 The system is the first to exhibits superconductivity as a low dimensional vanadate.
 Superconductivity exhibits a bell-shaped $T_{c}$ dependence with an optimum $T_{c}$
 of about 7K at 8GPa on the \emph{P-T} phase diagram. The cations (A=Li,Na,Ag) occupy
 one of the two nearest-neighboring sites. The cation disorder-order
 transition occurs in the metallic phase.$^{1, 9, 10}$ The cation-,
 charge-, and spin-ordering temperatures, $T_{A}$, $T_{co}$, and $T_{AF}$, respectively, are listed in Table I.
 As for $\beta$-Ag$_{0.33}$V$_2$O$_5$, $T_{A}$, $T_{co}$, and $T_{AF}$ are 200,
 90, and 27K, respectively.

\begin{table}
\caption{\label{tab:table1}Cation-, charge-, and spin-ordering
temperatures, T$_{A}$, T$_{CO}$, and T$_{AF}$ at ambient pressure.
In $\beta$-Sr$_{0.33}$V$_2$O$_5$ , the cation disorder-order
transition already occurs at room temperature, and the ground state
of insulating phase is not an antiferromagnetic states, but a
spin-gapped state.$^{7, 8, 10-12}$}
\begin{ruledtabular}
\begin{tabular}{ccccc}
 compound & V$^{4+}$/V$^{5+}$ & $T$$_{A}$ & $T$$_{CO}$ (K) & $T$$_{AF}$ \\ \hline
 $\beta$-Li$_{0.33}$V$_{2}$O$_{5}$ & 1/5 & - & 180 & 7 \\
 $\beta$-Na$_{0.33}$V$_{2}$O$_{5}$ & 1/5 & 260 & 136 & 24 \\
 $\beta$-Ag$_{0.33}$V$_{2}$O$_{5}$ & 1/5 & 200 & 90 & 27 \\
 $\beta$-Sr$_{0.33}$V$_{2}$O$_{5}$ & 2/4 & $>$RT & 170 & - \\
\end{tabular}
\end{ruledtabular}
\end{table}

 Since the discovery of superconductivity, no
 further experiments have been performed other than \emph{P}
 dependence of $T_{c}$ because of experimental difficulties under
 high pressures. There still remain two major problems; one is the
 pairing symmetry and the other is behaviors near the phase boundary.
 The detailed boundary has not still been fixed; one possibility is phase
 separation between the CO and SC phases, and the other is crossover
 between them. In relation to this problem, a novel mechanism such as
 a charge-fluctuation mediated mechanism has been proposed as the
 superconducting mechanism.$^{13}$

 Besides superconductivity, the insulating phase including the CO and AF phases, caused by
 metal-insulator transition (MIT) at low pressures below 7GPa, also exhibits
 unique features. The successive charge and magnetic orderings are realized although
 the electron density is very low; a 3d$^1$ electron exists among six V sites as
 is formally expressed as  $\beta$-A$_{0.33}$(V$^{4+}$+5V$^{5+}$)$_{0.33}$O$_5$, where V$^{4+}$ is a magnetic ion with spin
 S=1/2 and V$^{5+}$ is a nonmagnetic ion. Several charge-distribution models have been proposed so far since the discovery
 of superconductivity. Appearance of various models comes from the complicated
 crystal structure; there exist three crystalographically inequivalent Vi sites, V1, V2, and V3. Each
 of them forms quasi-one-dimensional coupling along b axis.

 $\beta$-Na$_{0.33}$V$_2$O$_5$, a prototype of the series, has been investigated so far using several experimental techniques.
 Yamaura \emph{et. al.} observed the lattice modulation along the b direction with twofold (2b) and sixfold (6b) periodicity below $T_{A}$ and
 $T_{CO}$, respectively, and proposed a rectangle-type charge-ordering model below $T_{CO}$ from the X-ray analysis, where V$^{4+}$ ions are
 located on every three V2 sites in the b axis.$^{10,14}$
 Nagai \emph{et. al.} also observed from the neutron diffraction the same lattice modulations along the b direction.$^{15}$
 Moreover, they suggested the charge disproportionation with 3b periodicity from the analysis of
 magnetic Bragg reflections, and attributed the 6b lattice modulation below $T_{CO}$ to the
 2b lattice modulation multiplied by the 3b charge distribution.
 They proposed a charge model which includes a nonmagnetic V site in every three Vi sites along the b axis.
 Relative charge density for the Vi sites was estimated as 3:2:3, respectively.
 Meanwhile electron spin resonance (ESR) suggested another model; the electrons
 are primarily located on the V1 zigzag chains and the charges occupy six consecutive V1
 sites within the V1 zigzag chain below $T_{CO}$. $^{16}$
 $^{51}$V- and $^{23}$Na-NMR measurements on $\beta$-Na$_{0.33}$V$_2$O$_5$
 were investigated based on the charge-distribution model proposed by the neutron
 diffraction measurement and relative charge density for the Vi sites was estimated as 3:1:1.$^{17, 18}$
 The NMR results suggest that V1 site possesses the most charge,
 whereas charge distribution is rather uniform according to the neutron diffraction.
 They also reported the changes of nuclear spin-lattice relaxation rate $1/T_{1}$ at every transition temperatures. $^{19}$

 $\beta$-Ag$_{0.33}$V$_2$O$_5$, an isomorphic compound of $\beta$-Na$_{0.33}$V$_2$O$_5$, has an advantage over $\beta$-Na$_{0.33}$V$_2$O$_5$
 because crystallographically stable single phase is available for $\beta$-Ag$_{0.33}$V$_2$O$_5$ whereas $\beta$-Na$_{0.33}$V$_2$O$_5$ is very sensitive to Na concentration;
 $\beta$-Na$_{0.32}$V$_2$O$_5$ hardly exhibits metallic behavior in the whole $\emph{T}$ range, although $\beta$-Na$_{0.33}$V$_2$O$_5$ is
 metallic at room temperature.$^{8}$
 In $\beta$-Ag$_{0.33}$V$_2$O$_5$ we proposed a much simple model in the previous NMR and ZFR works; a 3d$^1$ electron is located on a rung of V1-V3 and
 V2-V2 ladders like a proton molecular orbital. The electric quadrupole frequency and internal field which correspond to a half of a 3d$^1$ electron
 were observed in the CO and AF phases, respectively. These results give an evidence of the charge sharing. $^{20}$

 $\beta$-Sr$_{0.33}$V$_2$O$_5$, which possesses more charge than $\beta$-Na$_{0.33}$V$_2$O$_5$ or
 $\beta$-Ag$_{0.33}$V$_2$O$_5$, exhibits a similar charge ordering,
 although ground state properties are different. $^{6, 10, 21-23}$
 The $^{51}$V-NMR measurements were performed in
 the metallic phase and remarkable charge disproportion was reported in the V2-V2 ladders;
 development of ferromagnetic correlation at low temperatures was suggested from
 the analysis of Knight shift and $1/(T_{1}T)$. $^{24}$

 At present a variety of phenomena were observed so far for $\beta$-vanadium bronzes. It is not clear whether the variety
 is systematically explained from theoretical aspects. Nuclear
 magnetic resonance (NMR) measurement is one of the most powerful methods
 for the systematical understanding because the method is site
 selective, namely can specify microscopic information of 3d$^1$ electrons
 on each Vi site independently. In the previous work, we proposed a
 model of the CO phase in $\beta$-Ag$_{0.33}$V$_2$O$_5$. In this paper, we extended the previous NMR work in detail focusing
 on the electric field gradient (EFG) and spin ordering.

\section{CRYSTAL STRUCTURE}
 The series of compound consists of three kinds of inequivalent Vi sites.
 As shown in Fig. 1 a complicated crystal structure was often used in
 the studies of an early stage; the edge-shared zigzag chains of V1O$_{6}$ octahedra,
 the corner-shared two-leg ladders of V2O$_{6}$ octahedra, and the edge-shared zigzag chains of V3O$_5$ pyramids. $^{10, 21, 25}$

\begin{figure}
\includegraphics[width=7.8cm]{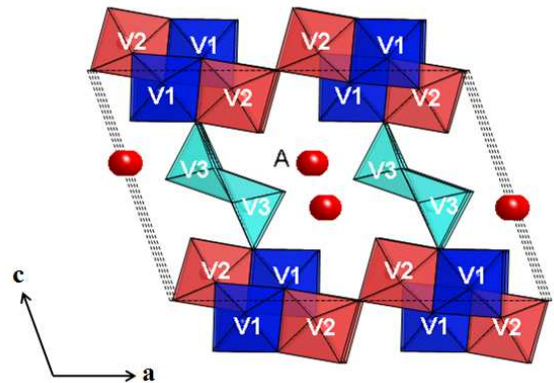}
\caption{\label{fig:epsart} Crystal structure of
$\beta$-A$_{0.33}$V$_2$O$_5$ (A=Li,Na,Ag) series. V1O$_{6}$
octahedra and V3O$_{5}$ pyramids form edge-shared zigzag chains and
V2O$_{6}$ octahedra form corner-shared two-leg ladders along the b
axis. Cations occupy one of two nearest-neighboring sites in the
same ac plane. $^{10, 21, 25}$}
\end{figure}

 However, a much simpler theoretical model has been presented recently by
 Doublet and Lepetit.$^{26}$ Fig. 2 shows weakly interacting two-leg ladder model based on
 extended H\"{u}ckel tight-binding calculations; the V1O$_5$ and V3O$_5$ pyramids form
 corner-shared two-leg ladders, and the two V2O$_5$ pyramids also form another
 ones independently. Cations are located in the tunnels of the framework
 and occupy one of two nearest-neighboring sites as shown in both Figs. 1 and 2.

\begin{figure}
\includegraphics[width=7.8cm]{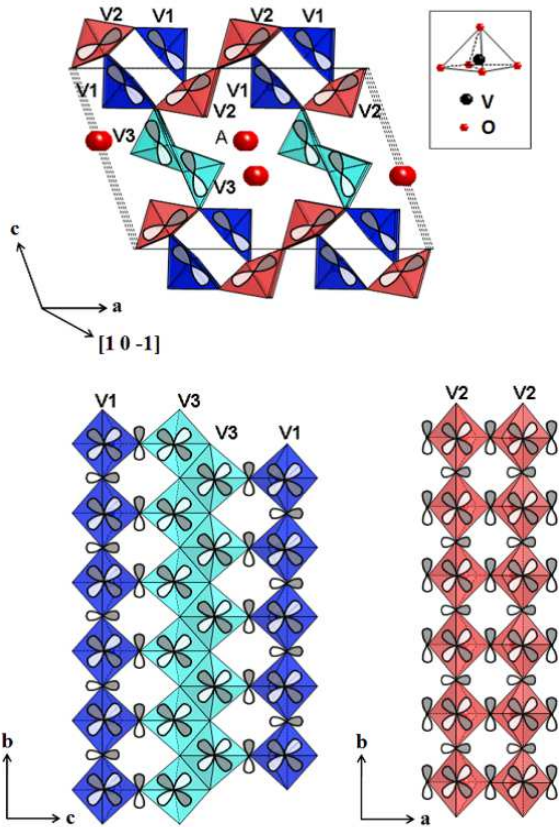}
\caption{\label{fig:epsart} Two-leg ladder model proposed by Doublet
and Lepetit.
 The covalent bonds of V orbital form two-leg ladders. V1 and V3 sites form two-leg ladders and two V2s sites form
  the other independent ones. The inset represents VO$_5$ pyramid. $^{26}$}
\end{figure}

 The crystalographical symmetry  of the uniform metallic phase above $T_{A}$ is C2/m.
 Because of the zigzag ordering of the cations, this symmetry changes to $P2_{1}/a$ and the b cell parameter becomes twice as large as that above $T_{A}$.
 Although it retains the same space group below $T_{CO}$, the b cell parameter becomes six times as large as that above $T_{A}$. $^{10, 14, 21}$
 Thus, the sixfold superstructure is deeply concerned with the charge ordering.

\section{EXPERIMENTAL CONDITIONS}
 $^{51}$V-NMR measurements were performed using a  needlelike $\beta$-Ag$_{0.33}$V$_2$O$_5$ single
 crystal prepared by a self-flux method with a CZ furnace. The crystal grew up
 in the direction of the b axis, the chain or leg direction. The volume of the
 crystal was 0.5$\times$4$\times$0.5 mm$^3$. We used a conventional pulsed-NMR spectrometer.
 NMR spectra were measured at a frequency of 59.00MHz for the field ($\textbf{\emph{H}}$ ) parallel (${\parallel}$)
 or perpendicular (${\perp}$) to the b axis. We attached a hand-made rotation apparatus to
 the NMR probe to rotate sample for the measurement with $\textbf{\emph{H}}$${\perp}$b.
 Hereafter we express the direction of the $\textbf{\emph{H}}$ as an angle $\theta$ from the
 [1 0 -1] direction as shown in Fig. 2.  The error margin was within \
 $\pm$ 5 degrees caused in an initial setting. A powder pellet sample
 of 8-mm diameter and 6-mm length was used in Zero Field Resonance (ZFR) measurement.
 ZFR spectra were measured up to 170 MHz with a separation
 of 0.25 MHz / point at 4.2K.

\section{EXPERIMENTAL RESULTS}
 \subsection{$^{51}$V-NMR for \emph{\textbf{H}}${\parallel}$b}

 We measured $^{51}$V-NMR spectra for
 \emph{\textbf{H}}${\parallel}$b in the $T$ range, 4.2K$<$$T$$<$300K.
 The pattern of $^{51}$V(I=7/2)-NMR spectra changed at 200, 90, and
 27K corresponding to the transition temperatures $T_{Ag}$,
 $T_{CO}$,and $T_{AF}$, respectively. The typical
 $\textbf{\emph{H}}$-swept spectra in the metallic, CO, and AF phases
 are presented in Fig. 3.
 The spectra in the metallic phase were measured at both
 $T$$>$$T_{A}$ and $T$$<$$T_{A}$.
 There appear seven peaks with a
 constant separation for a V site owning to the nuclear quadrupole
 interaction. We observed three sets of signals originating from
 three crystallographically inequivalent V sites above $T_{AF}$.
 The sites were directly assigned from the data for
 $\textbf{\emph{H}}$${\perp}$b. Information of the peak separations
 appearing in Fig. 3 is reflected in the fitting curves of Figs.
 6(b)-8(b) in Section IV B via asymmetric factors of the EFG. The site
 assignment in Fig. 3 were derived from the analysis.

\begin{figure}
\includegraphics[width=8.0cm]{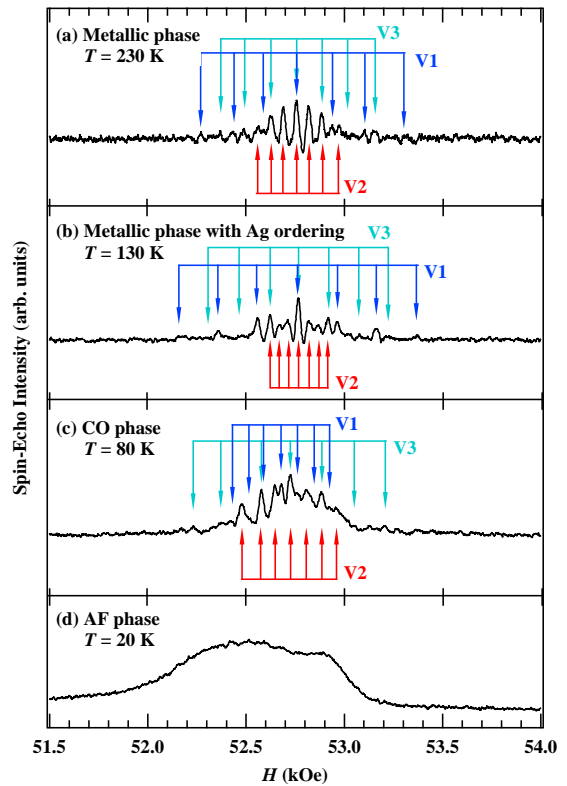}
\caption{\label{fig:epsart} $^{51}$V-NMR spectra for
\emph{\textbf{H}}${\parallel}$b in the metallic, CO, and AF phases.
Disorder-order transition of Ag$^+$ occurs at 200K in the metallic
phase.}
\end{figure}

 In the CO phase, there appeared broad basal signals in addition to the
 sharp peaks. The sharp peaks come from magnetic V$^{4+}$
 like sites, while the broad signals from nonmagnetic V$^{5+}$ sites.
 As shown in Fig.4, the relative intensity of the sharp peaks becomes
 small compared to the broad signals with increasing $\tau$, separation of two rf pulses.
 This implies that the spin-spin relaxation time $T_{2}$ of the sharp peaks is shorter
 than that of the broad signals. In general, nonmagnetic site possesses much larger
 $T_{2}$ than that of the magnetic site. Therefore, the basal signals are
 attributed to nonmagnetic sites. In fact, we measured $\tau$
 dependence of the spin echo for the first satellite. The
 position for the V1 sites is shown in Fig. 4 as bold-type arrows on a
 dashed line. The $\tau$ dependence is shown in Figs. 5(a) and 5(b).
 Fig. 5(b) is an expansion of Fig. 5(a). The spin-echo modulation due
 to the nuclear-quadrupole interaction was clearly seen at short
 $\tau$ region in the decay curves. We also plotted the $\tau$
 dependence in the metallic phase for comparison. The decay curve in
 the CO phase consists of two components with shorter and longer
 characteristic time constants. The decay curve with a shorter time
 constant is similar to that in the metallic state. Therefore the
 shorter component is attributed to the magnetic sites. The results
 of NMR spectra and the $\tau$ dependence of the spin echo are
 consistent.

\begin{figure}
\includegraphics[width=7.8cm]{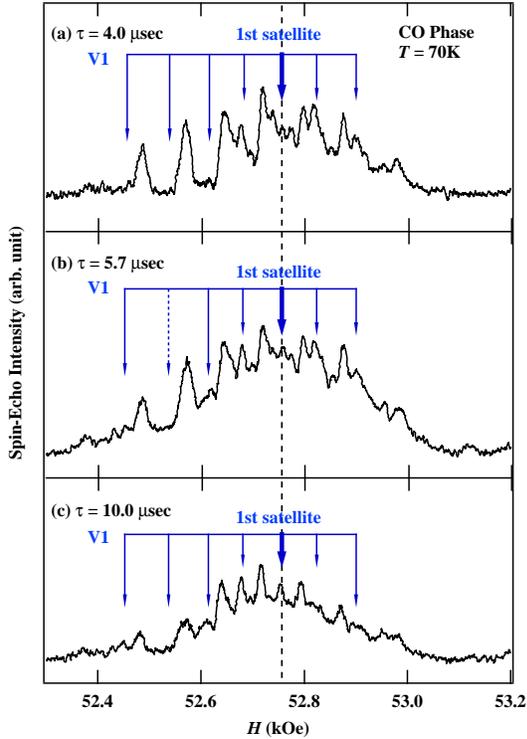}
\caption{\label{fig:epsart} $\tau$ dependence of $^{51}$V-NMR
spectra for \emph{\textbf{H}}${\parallel}$b in the CO phase.
Relative intensity of basal broad signals increases with increasing
$\tau$. Echo-decay curves in Figs. 5(a) and 5(b) were measured at
the position shown by bold-type arrows.}
\end{figure}

\begin{figure}
\includegraphics[height=13cm]{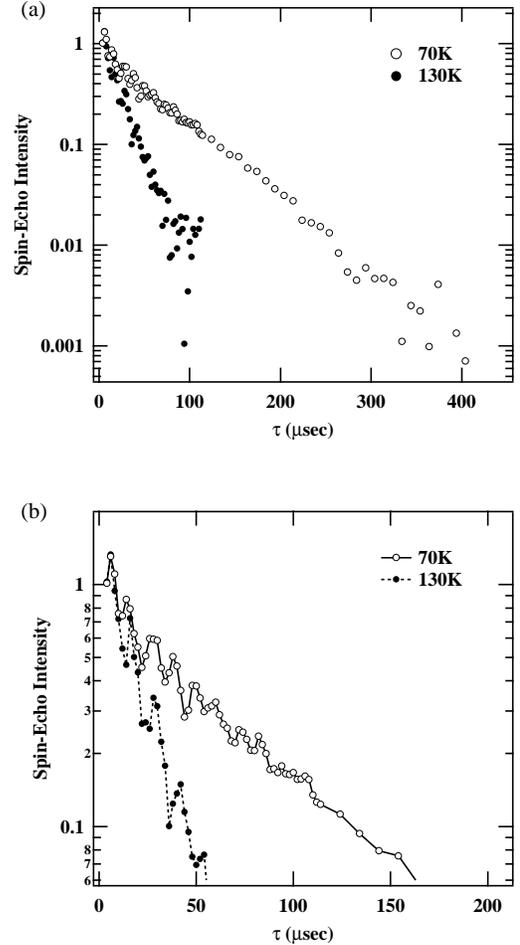}
\caption{\label{fig:epsart} Pulse separation ($\tau$) dependence of
the spin-echo intensities in the CO and the metallic phases. Open
and closed circles represent the decay curve in the CO phase and
metallic phase, respectively. (a)Decay curve in the CO phase
includes two components. The shorter one is similar to that in the
metallic phase. (b) An expansion of Fig. 5 (a). Modulation due to
the nuclear quadrupole interaction was observed in both phases.}
\end{figure}

 In the AF phase, the sharp signals arising from magnetic sites are wiped
 out and only broad signals were observed.
 It is because the signals of magnetic sites were shifted into higher frequencies
 owning to the large internal field caused by the AF magnetic ordering.

 \subsection{$^{51}$V NMR for \emph{\textbf{H}}${\perp}$b and Site Assignment}

\begin{figure}
\includegraphics[height=15.8cm]{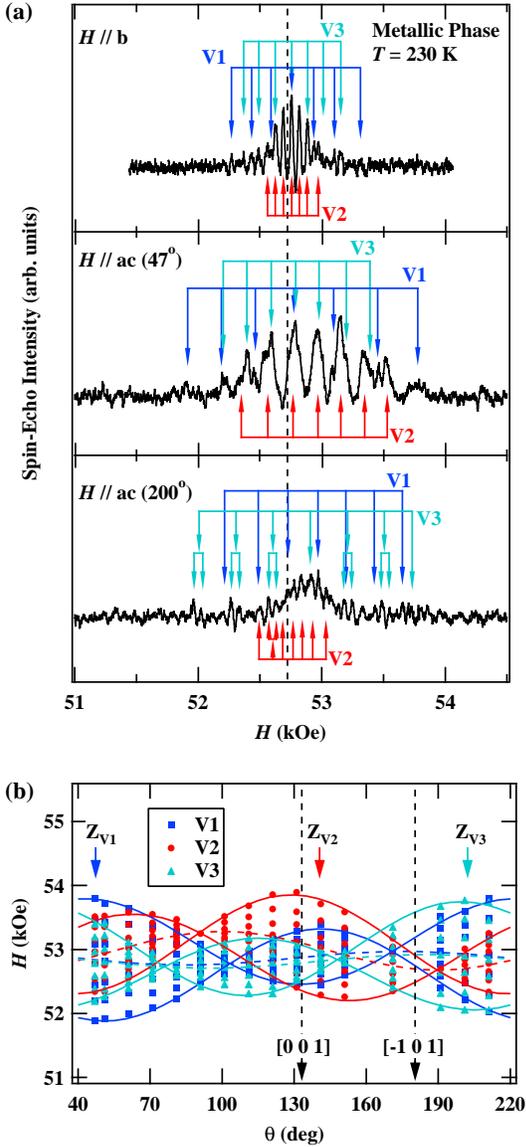}
\caption{\label{fig:epsart} (a) $^{51}$V-NMR spectra in the metallic
phase at 230K. (b) $\theta$ dependence of the $^{51}$V-NMR peaks.
The third satellite peaks and Knight shifts are indicated by solid
curves and dotted curves, respectively. Each Z$_{Vi}$ represents the
angle of the first principal axis of EFG tensor for each Vi sites.}
\end{figure}

\begin{figure}
\includegraphics[height=15.8cm]{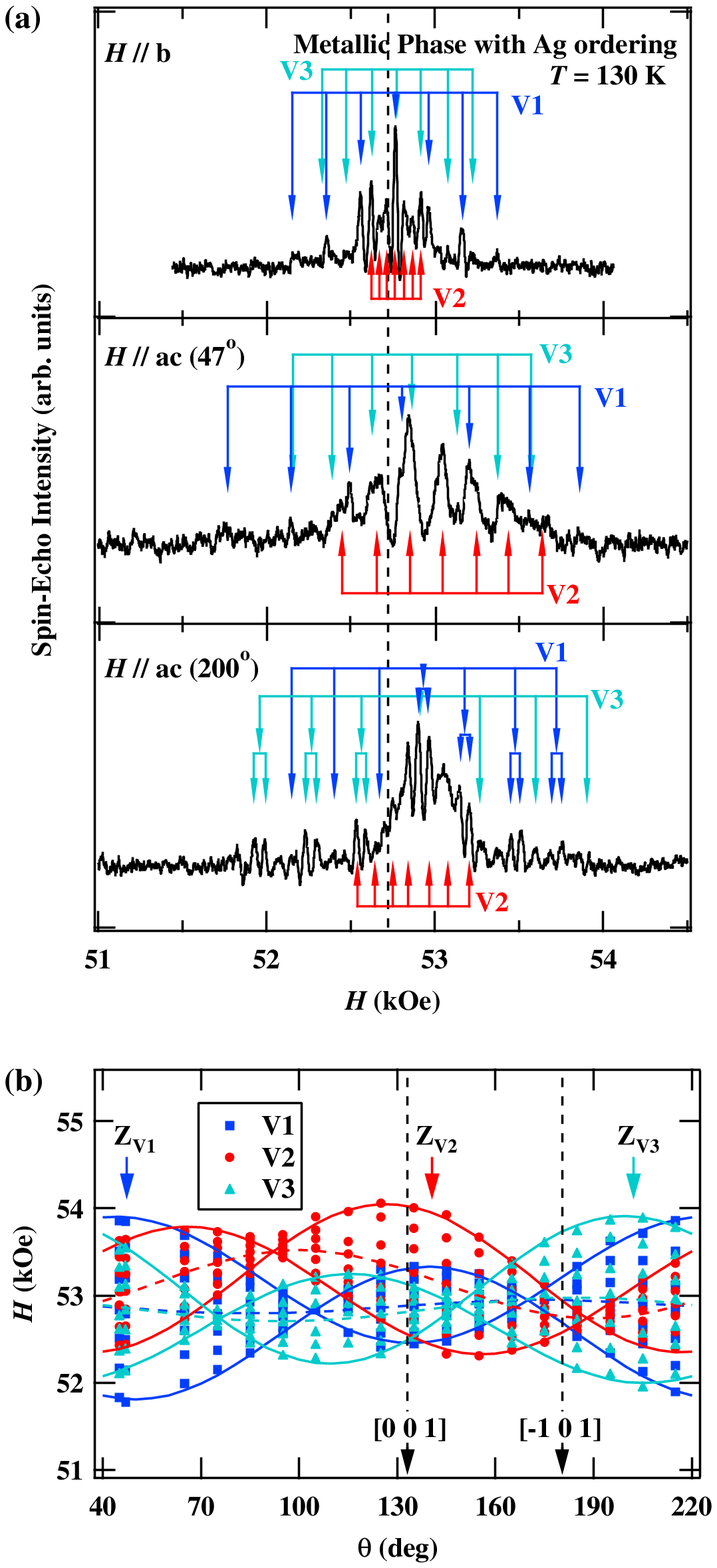}
\caption{\label{fig:epsart} (a) $^{51}$V-NMR spectra in the metallic
phase with Ag ordering at 130K. (b) $\theta$ dependence of the
$^{51}$V-NMR peaks. The third satellite peaks and Knight shifts are
indicated by solid curves and dotted curves, respectively. Each
Z$_{Vi}$ represents the angle of the first principal axis of EFG
tensor for each Vi site.}
\end{figure}

\begin{figure}
\includegraphics[height=15.8cm]{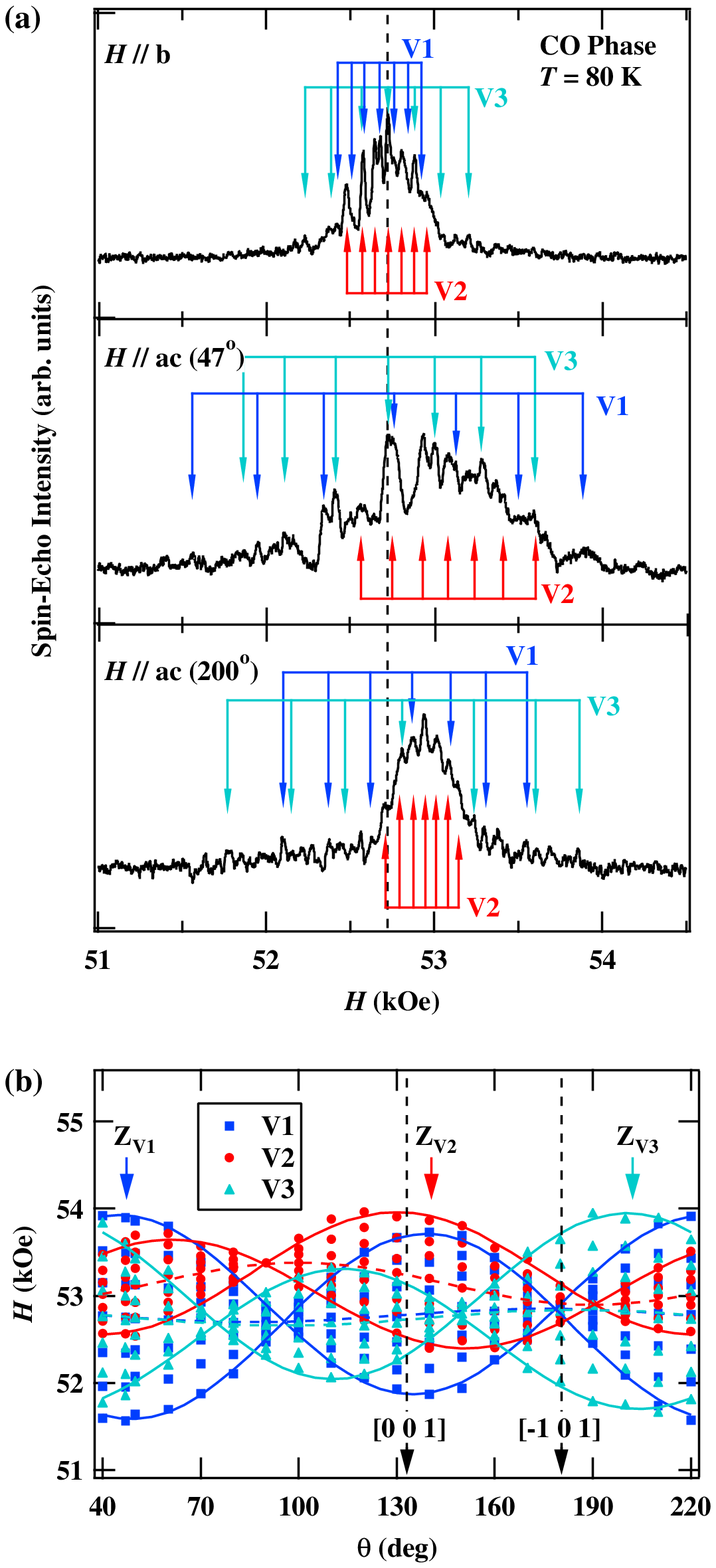}
\caption{\label{fig:epsart} (a) $^{51}$V-NMR spectra in the CO phase
at 80K. (b) $\theta$ dependence of the $^{51}$V-NMR peaks. The third
satellite peaks and Knight shifts are indicated by solid curves and
dotted curves, respectively. Each Z$_{Vi}$ represents the angle of
the first principal axis of EFG tensor for each Vi site. The data
were published in Ref. 20.}
\end{figure}

 We measured $^{51}$V-NMR spectra for the $\textbf{\emph{H}}$ direction of
 $\theta$=47$^\circ$ or 200$^\circ$ in the $T$ range, $T$$>$$T_{AF}$.
 The typical spectra at 230, 130, and 80K are shown in Figs. 6(a), 7(a),and 8(a),respectively,
 as well as those for $\textbf{\emph{H}}$${\parallel}$b.
 These angles correspond to directions of the principle axes of EFG tensor for three Vi sites;
 the direction of $\theta$=47$^\circ$ corresponds to the first and
 second principle axes of the V1 and V2 sites, respectively, whereas that
 of $\theta$=200$^\circ$ corresponds to the first principle axis of the V3 sites.
 The directions of the principle axes of the EFG tensor are determined from the following analysis in section V.
 As seen in Fig. 8(a), there appeared the broad basal signals for
 \emph{\textbf{H}}${\perp}$b, the same as that for $\textbf{\emph{H}}$${\parallel}$b.

 Some peaks at $\theta$=200$^\circ$ seem to split into two in the
 metallic phase for the first look. These splittings were also seen
 in the CO phase. Because these splittings are seen in the uniform
 metallic phase, it comes not from the splitting due to the cation
 ordering or charge ordering but from a large free induction decay
 (FID) overlapped with the spin echo. In such a case we estimated
 the resonance field taking mean values of the two positions.

 We also measured $\theta$ dependence of the $^{51}$V-NMR peaks at 230, 130,
 and 80K, as shown in Figs. 6(b), 7(b), and 8(b), respectively.
 The dotted and two solid lines in these figures indicate fitting
 curves for Knight shifts and the third satellite peaks, respectively
 (See Eq. (1) and (2) in section V).
 The V1, V2, and V3 sites were assigned using the
 directions of the first principal axes of EFG tensor.
 According to the theoretical investigation, hybridization
 of V orbitals forms weakly interacting two-leg ladders as mentioned
 in Section II. The ground state of V orbital is expected to be
 d$_{xy}$, therefore the first principal axis should turn to the
 apical directions of each VO$_5$ pyramid. In other words, the
 splitting due to EFG should become the maximum when the
 $\textbf{\emph{H}}$ is applied to the apical direction. Our
 experimental results are consistent with the theoretical model,
 namely the directions of the first principal axes agree with the
 apical directions within the angles of 23, 7, and 24 degrees for the V1,
 V2, and V3 sites, respectively. Three sets of the seven peaks were
 identified at almost all angles, however, some peaks could not be
 assigned because of the overlap of large FID signals.


 The principle axes of the Knight shift, a direction where the central peak becomes the maximum, were estimated
 to be 175, 100, and 5$^\circ$ for the V1, V2, and V3 sites, respectively.
 In the case of d$_{xy}$ symmetry, the principle axes of the Knight shift should be the same with those of the EFG.
 However, the directions of the principle axes do not agree with each other.
 In other words, the first principle axes of the Knight shift do not
 coincide with the apical directions of the pyramids. The discrepancy
 would arise from distortion of the local VO$_5$ pyramids. Thus the
 Knight shift offers less information than the EFG for the site assignment.

 \subsection{Zero Field Resonance in AF Phase}
 The signals originating from the magnetic V sites are not detected from NMR
 measurements in the AF phase. They were shifted into high frequency because of
 the large internal field caused by the magnetic ordering. We preformed
 the Zero-field resonance (ZFR) at 4.2K to investigate the spin alignment
 in the AF phase. Figure 9 shows the ZFR spectra observed up to 170MHz.
 Three coils were used to cover whole frequency
 region, and the intensity were adjusted to overlap smoothly at the boundary
 of frequency span which one coil can cover.
 Three sets of pairs with frequency separation of 50MHz were observed
 except for the peaks at 83.5MHz. The relative intensity differs
 between two peaks in each pair; The intensity observed at higher
 frequencies is larger than that at lower frequencies. The difference
 does not reflect the number of magnetic V sites contributing to the
 peaks because the resonance condition is not the same throughout the
 frequency region. Detail of the site assignment was mentioned in
 the following section. The peaks at 21.50MHz and 71.25MHz
 were assigned as peaks originating from the V1 sites, 48.00MHz and
 98.75MHz from the V2 sites, and 24.75MHz and 73.00MHz from the V3
 sites. At present, the origin of 83.50MHz is not certain.

\begin{figure}
\includegraphics[width=8.0cm]{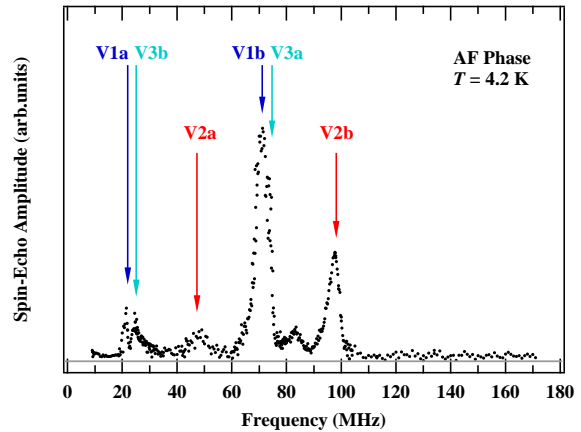}
\caption{\label{fig:epsart} Zero-field resonance spectrum up to
170MHz. The data were published in Ref. 20.}
\end{figure}

\section{ANALYSIS }
 \subsection{Analysis of NMR results}
 \subsubsection{Fitting formula}
 Seven peaks with an equal frequency separation originate from the
 interaction of nuclear quadrupole moment Q with the EFG experienced
 by the nucleus. When $\theta$-$\theta$$_{0}$ is denoted as an angle between the first principle axis and the
 $\textbf{\emph{H}}$, and $\phi$-$\phi$$_{0}$  as an angle between the
 third principal axis and the $\textbf{\emph{H}}$, the frequency separation for an arbitrary
 $H$ direction is expressed as
\begin{eqnarray}
  {\nu}^{(1)}_{m \rightarrow m-1} =
  \frac{\nu_{Q}}{2} (m-\frac{1}{2})
  \{3\cos ^2 (\theta - \theta _0)-1  \notag \\
  -\eta \sin ^2 (\theta - \theta _0)\cos 2 (\phi - \phi _0)\} \notag \\
  \notag \\
  (m=i+1/2, \vert i \vert \leq 3)
\end{eqnarray}
 where $  {\nu}_{Q} =\dfrac{3e^{2}qQ}{h2I(2I-1)} \notag $ is the electric quadrupole frequency
 and $  q =\dfrac{\partial^2 V}{\partial z^2} (=V _{ZZ}) $ is the EFG
 for the first principle axis. The asymmetric parameter $\eta$ is
 defined as $ \eta= \dfrac{V _{XX} - V _{YY}}{V _{ZZ}} $ using diagonal elements of the EFG tensor. In the case that the
 $\textbf{\emph{H}}$ directs the first principle axis, the frequency span
 between the two first satellites is given as 2$\nu _Q$. The intensities of seven peaks
 are proportional to the transition ratio
 $ \vert<I m|I ^+|I m-1>\vert^2 $ , therefore the ratio should be 7:12:15:16:15:12:7
 with increasing $m=i + 1/2$ $(|i|\leq 3)$.

 The $^{51}$V Knight Shift $\emph{K}$ is defined as
 $ K \equiv \dfrac {H _{ref} - H _{0}}{H_{0}}$
 where $H _{0}$ represents the field of the central peaks for the V1, V2, and V3 sites.
 The reference field $H_{ref}$ is calculated as 52.71 kOe from the frequency of 59.00 MHz and
 the gyromagnetic ratio $\gamma_N$= 1.1193 MHz/kOe of $^{51}$V. The
 angle dependence of the Knight shift is caused by the asymmetry of
 hyperfine field:
\begin{eqnarray}
  {K}(\theta) = {K}_{l} + K_{a}\{3 \cos^2(\theta - \theta_{K}) -1\}
\end{eqnarray}
 where ${K}_{l}$ and ${K}_{a}$ are the symmetric and asymmetric terms, respectively. $\theta$-$\theta_{K}$ represents the angle
 between the symmetry axis and the $\textbf{\emph{H}}$.
 The parameters ${\nu}_{Q}$, $\eta$, ${K}_{l}$, ${K}_{a}$ and ${\theta}_{K}$ at 230, 130, and 80K are
 estimated from the $^{51}$V-NMR results with the $\textbf{\emph{H}}$
 parallel to the ac plane (Figs. 6(b)-8(b)).
 The $\theta$ dependence of the peak positions is analyzed based on Eqs. (1) and (2).
 We used the definition of the EFG tensor,
 $\vert V_{XX} \vert \leq \vert V_{YY} \vert \leq \vert V_{ZZ} \vert$
 and $ V_{XX} + V_{YY} + V_{ZZ} = 0$ in the analysis. The value of ${\nu}_{Q}$, $\eta$,
 ${K}_{l}$, ${K}_{a}$ are listed in the Tables II, III and IV.

\begin{table}

 \caption{\label{tab:table1}Parameters in the metallic phase.}
 \begin{ruledtabular}
 \begin{tabular}{cccccc}
 230K & Site & ${\nu}_{Q}$(MHz) & $\eta$ & ${K}_{l}$(\%) & ${K}_{a}$(\%) \\ \hline
   & V1 & 0.35 & 0.10 & -0.22 & -0.13 \\
   & V2 & 0.28 & 0.46 & -0.32 & -0.37 \\
   & V3 & 0.31 & 0.06 & -0.13 & -0.13 \\
 \end{tabular}
 \end{ruledtabular}

 \caption{\label{tab:table1}Parameters in the metallic phase with Ag ordering.}
 \begin{ruledtabular}
 \begin{tabular}{cccccc}
 130K & Site & ${\nu}_{Q}$(MHz) & $\eta$ & ${K}_{l}$(\%) & ${K}_{a}$(\%) \\ \hline
   & V1 & 0.39 & 0.17 & -0.26 & -0.09 \\
   & V2 & 0.28 & 0.63 & -0.54 & -0.48 \\
   & V3 & 0.36 & 0.07 & -0.17 & -0.17 \\
 \end{tabular}
 \end{ruledtabular}

 \caption{\label{tab:table1}Parameters in the CO phase.}
 \begin{ruledtabular}
 \begin{tabular}{cccccc}
 80K & Site & ${\nu}_{Q}$(MHz) & $\eta$ & ${K}_{l}$(\%) & ${K}_{a}$(\%) \\ \hline
   & V1 & 0.44 & 0.57 & -0.07 & -0.09 \\
   & V2 & 0.27 & 0.36 & -0.66 & -0.30 \\
   & V3 & 0.42 & 0.13 & -0.02 & -0.11 \\
 \end{tabular}
 \end{ruledtabular}

\end{table}

 The values of ${\nu}_{Q}$ for the V1 sites is close to that for the
 V3 sites at each temperature; they are about 0.3MHz at 230K, and
 become larger at 80K, namely 0.44 and 0.42MHz for the V1 and V3
 sites, respectively.
 On the other hand, ${\nu}_{Q}$ for the V2 sites seems almost temperature
 independent and the value is about 0.28MHz.
 As for $\eta$, the value for the V1 sites is rather close to
 that for the V3 sites at 230K, however, there appears discrepancy
 between them at low temperature; $\eta$ for the V1 sites is more
 than four times as large as that for the V3 sites at 80K.
 The values of $\eta$ for the V2 sites is more than three times
 as large as that for the V1 and V3 sites in the metallic phase,
 but it decreases to less than that for V1 sites at 80K.

 The symmetric term ${K}_{l}$ of the Knight shift is negative for all Vi
 sites. The absolute values for the V1 and V3 sites in the metallic
 phase are larger than those in the CO phase. On the other hand, the
 absolute value for the V2 sites gradually increases with decreasing
 temperature, and is several times as large as those for the V1 and
 V3 sites.
 The value of ${K}_{a}$ for the V1 sites is almost the same with that
 for the V3 sites at 230K. They hardly change even in the CO phase.
 On the other hand, the value for the V2 sites is three times as
 large as those for the V1 and V3 sites at 230K.

 These results show that the V1 sites are in a similar local environment to the V3 sites,
 suggesting covalent bonds between the V1 and V3 sites. On the other
 hand, the V2 sites seem independent of the V1 and V3 sites. This
 analysis is consistent with the model of Doublet and Lepetit.
 Moreover, we mention the charge distribution in the CO phase.
 ${\nu}_{Q}$ is determined by the charge density of the on-site V
 because contribution from the surrounding oxygen sites is small.
 The fact is easily seen by comparing ${\nu}_{Q}$ for magnetic
 V$^{4+}$O$_5$ pyramids and nonmagnetic V$^{5+}$O$_5$ pyramids:
 ${\nu}_{Q}$ observed in CaV$_2$O$_5$ and V$_2$O$_5$, are 0.97 and
 0.06 MHz, respectively.$^{27}$ In $\beta$-Ag$_{0.33}$V$_2$O$_5$ the
 values of ${\nu}_{Q}$ for the three V sites have the intermediate
 values (0.27 to 0.44 MHz) in the CO phase, suggesting that each
 magnetic site possesses low charge density of 1/2 to 1/4. The ratio
 of the weight is estimated as 5:3:5 for the V1, V2 and V3 sites,
 respectively. The facts suggest that the charge distribution is
 roughly equal for the three V sites, respectively. To maintain the
 electrical neutrality, a 3d$^1$ electron exists per six V sites,
 therefore the present results implies that a 3d$^1$ electron is
 shared with 2 to 4 sites and the other sites should be nonmagnetic.

 \subsubsection{$T$ dependence of the EFG tensor}

\begin{figure}
\includegraphics[height=10cm]{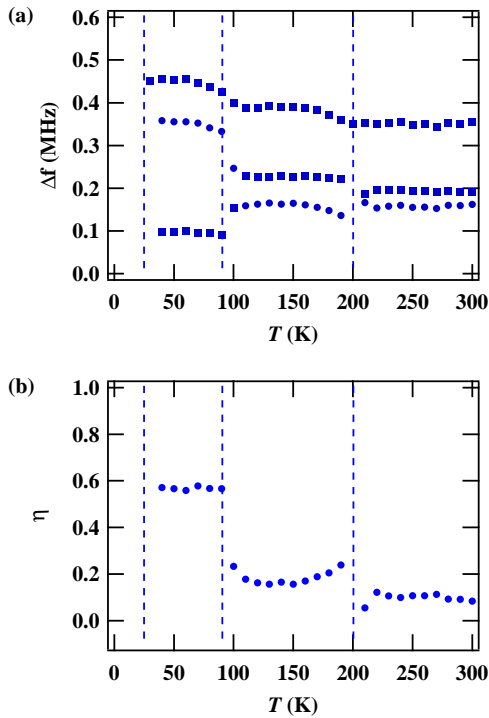}
\caption{\label{fig:epsart} (a)\emph{T} dependence of the peak
separation for the V1 sites. (b)\emph{T} dependence of ${\eta}$ for
the V1 sites.}
\end{figure}

\begin{figure}
\includegraphics[height=10cm]{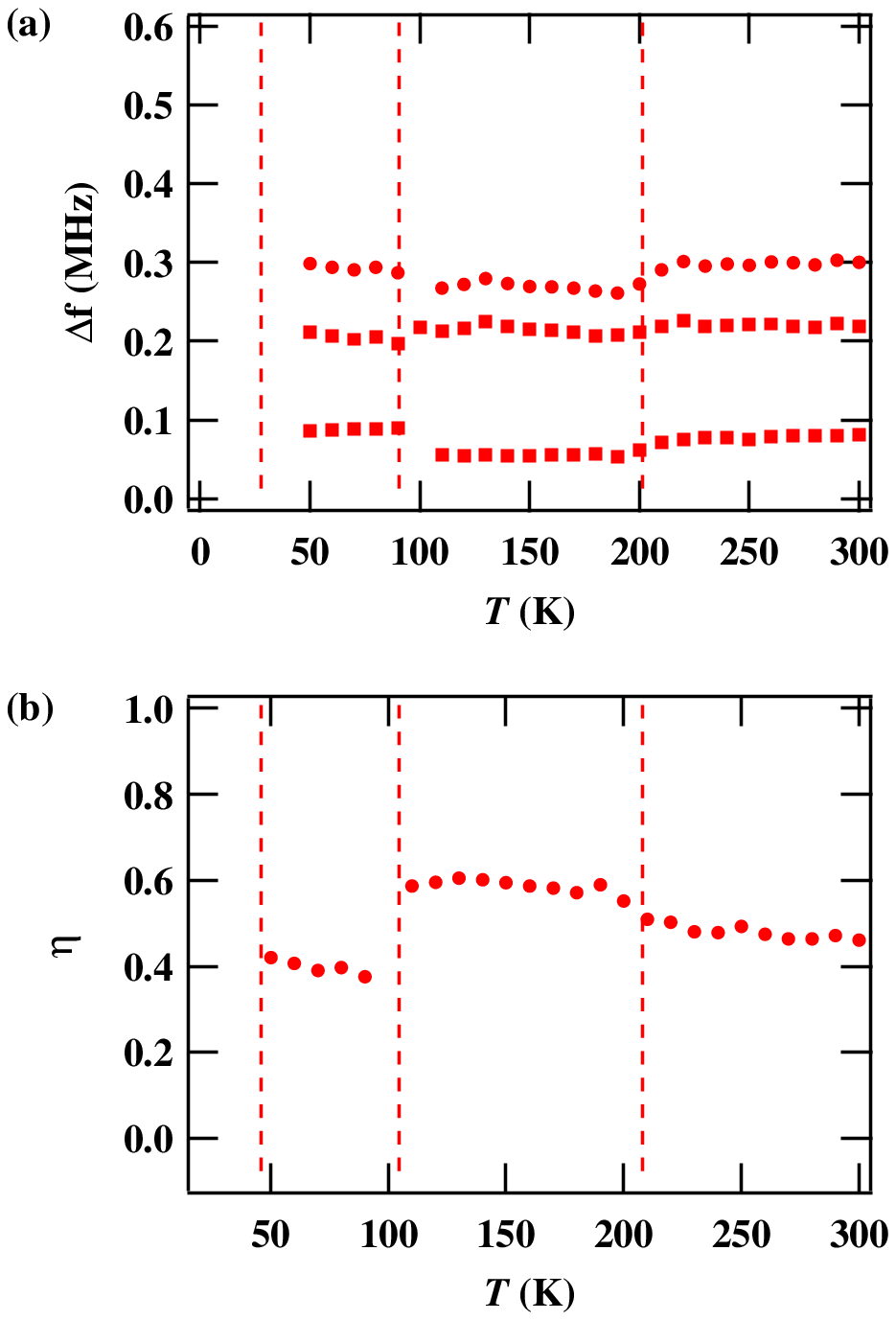}
\caption{\label{fig:epsart} (a)\emph{T} dependence of the peak
separation for the V2 sites. (b)\emph{T} dependence of ${\eta}$ for
the V2 sites.}
\end{figure}

\begin{figure}
\includegraphics[height=10cm]{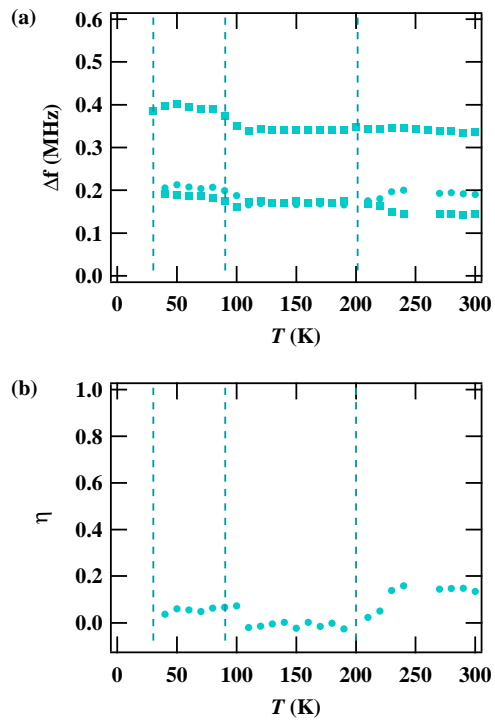}
\caption{\label{fig:epsart} (a)\emph{T} dependence of the peak
separation for the V3 sites. (b)\emph{T} dependence of ${\eta}$ for
the V3 sites.}
\end{figure}

 In the previous section, we have discussed the parameters in each phase.
 Here we present the $T$ dependence of the peak separations for three principal axes and $\eta$ for each Vi site.
 The maximum peak separation in Figs. 10(a)-12(a) at 230, 130, and 80K, corresponds to ${\nu}_{Q}$ in the Tables II, III and IV, respectively.
 Closed squares in Figs. 10(a)-12(a), respectively, represent the peak
 separation measured for two principal axes of each Vi site, namely two of
 ${\nu}^{(1)}_{m \rightarrow m-1}({\theta}-\theta_{0}=0^\circ)(={\nu}_{Q})$,
 ${\nu}^{(1)}_{m \rightarrow m-1}({\theta}-\theta_{0}=90^\circ, {\phi}-\phi_{0}=0^\circ)$,
 and ${\nu}^{(1)}_{m \rightarrow m-1}({\theta}-\theta_{0}=90^\circ, {\phi}-\phi_{0}=90^\circ)$.
 As mentioned above,
 the direction of the first principal axis of the V1 sites and that
 of the second principal axis of the V2 sites correspond to
 $\theta$=47$^\circ$, and that of the first principle axis of the V3 sites corresponds to
 $\theta$=200$^\circ$. Another principal axis of each Vi site corresponds to the b axis.
 Closed circles represent calculated values for the other
 principal axis using the equation $V_{XX} + V_{YY} + V_{ZZ}= 0$.
 Figs. 10(b), 11(b), and 12(b) show the $\emph{T}$ dependence of $\eta$ for V1, V2 and V3 sites, respectively.
 The EFG tensor exhibits some features for each Vi site.

 (i) The EFG tensor of the V1 sites shows unique behavior; The b axis, the
 third principal axis in the metallic phase, turns into the second
 principal axis in the CO phase (Fig. 10(a)). Simultaneously, the value of $\eta$
 also enhances below $T_{CO}$ (Fig. 10(b)), reflecting that the distortion of
 the lattice is caused in the CO phase.

 (ii) The EFG tensor of the V2 sites shows only slight change throughout the $\emph{T}$ range between 30K and 300K (Fig. 11).

 (iii) The EFG tensor of the V3 sites is the most isotropic, namely, $\eta$ is the smallest among the three V sites (Fig. 12(b)).
 The value of ${\nu}_{Q}$ for the V3 sites is almost constant in the metallic phase and becomes a little bit larger below $T_{CO}$
 for all principle axes.(Fig. 12(a))

 Comparing the three Vi sites, ${\nu}_{Q}$ for the V1 and V3 sites show the similar \emph{T} dependence,
 but that for V2 sites is almost independent of \emph{T}.
 On the other hand, $\eta$ and the symmetry of the EFG tensor show the different feature for each Vi site.

 \subsection{Analysis of ZFR spectrum}
 Based on the analysis above mentioned, the V1 and V3 sites
 seem to be in a similar local environment. This behavior is also
 seen in the ZFR spectrum in the AF phase; the peaks at 21.50 and
 24.75 MHz are located on very close positions as well as 71.25 and
 73.00 MHz, suggesting that they originate from the magnetic V1-V3
 rungs. We assigned the peaks at 21.50 and 71.25 MHz to the magnetic
 V1 sites and those at 24.75 MHz and 73.00 MHz to the magnetic V3
 sites. The other signals at 48.00 and 98.75 MHz are attributed to the two V2s
 sites.

 We discuss the local spin alignment in the AF phase using the ZFR
 frequencies. ZFR frequency $f_{peak}$ is proportional to the
 internal field $H_{n}$ at each V site, i.e. $f_{peak} = \gamma _{N}
 \vert H_{n} \vert$. The internal field $H_{n}$ is composed of
 Fermi-contact field $H_{F}$ and dipole field $H_{dip}$: $\vert H_{n}
 \vert = \vert H_{F} + H_{dip} \vert$, where we neglected the hyperfine
 field due to the spin-orbit interaction by considering the g value
 for the isomorphic $\beta$-Na$_{0.33}$V$_2$O$_5$ observed in the
 ESR measurements.$^{16}$
 $\vert H_{n} \vert$ is expressed using expectation value of spin at a V site $<S>$, and
 hyperfine coupling defined as a magnetic field arising from one Bore
 magnetron. When we denote hyperfine coupling of $H_{F}$ as
 $A_{F}$, and dipole coupling as $A^{\parallel}_{dip}$ and
 $A^{\perp}_{dip}$ for spin parallel and perpendicular to the
 EFG maximum, $ \vert H_{n} \vert $ is written as
\begin{eqnarray}
 \vert \textbf{\textit{H}}_n \vert = 2 k <S> \sqrt{{A^{\parallel}}^2
 \cos ^2 \theta _n + {A^{\perp}}^2 \sin ^2 \theta _n}
\end{eqnarray}
 where
 ${A}^{\parallel (\perp)}= {\textit{A}}_F+{A}^{\parallel(\perp)}_{dip}$,
 $\it{k}$ is a reduction factor due to the covalent effect with oxygen, and ${\theta}_{n}$ is the angle between spin
 and the EFG maximum. $A^{\parallel}_{dip}$ and $A^{\perp}_{dip}$ for the d$_{xy}$ orbital are expressed as
\begin{eqnarray}
 {A}^{\parallel}_{dip}=-\frac{4}{7}{\mu_B}{<r^{-3}>}
\end{eqnarray}
 \ \ \ \ \ \ \ \ \  \ \ \ \ \ \ \ \ \ \ \ \ \ and
\begin{eqnarray}
 {A}^{\perp}_{dip}=\frac{2}{7}{\mu_B}{<r^{-3}>}.
\end{eqnarray}
 We used $k$=0.8 and $A_{F}$=-100 kOe as typical values. The values of
 $A^{\parallel}_{dip}$ and $A^{\perp}_{dip}$ are calculated to be -130 and 65 kOe, respectively,
 using the reference value $<r^{-3}>$ = 3.684 atomic units.$^{28}$

 Three Vi sites possess almost the same charge density from the analysis of ${\nu}_{Q}$, therefore
 $<S>$ should be expressed as
\begin{eqnarray}
 <S>=\frac{1}{2n}.
\end{eqnarray}
 X-ray diffraction shows the sixfold periodicity in the leg direction below $T_{CO}$, implying that
 possible choice of $n$ is an integer from one to six.

\begin{figure}
\includegraphics[width=8.0cm]{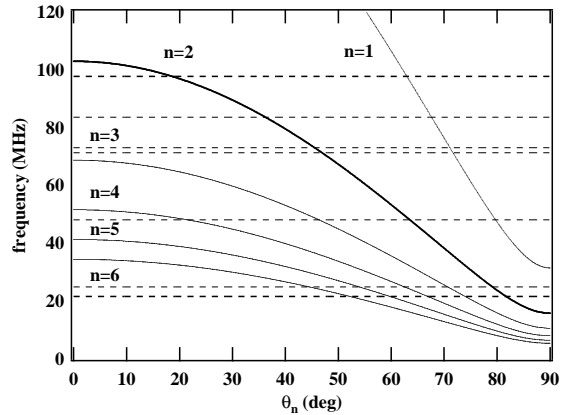}
\caption{\label{fig:epsart} ${\theta}_{n}$ dependence of the
resonance frequency for $n$=1 to 6. Dotted lines indicate resonance
frequencies of ZFR peaks.}
\end{figure}

 Fig. 13 shows the ${\theta}_{n}$ dependence of the resonance
 frequency calculated with Eqs. (3)-(6) for integers n=1 to 6.
 The signals below 24.75 MHz are out of the fitting curve for $n=1$,
 while the signals above 71.25 MHz are out of the fitting curve for $n \geq 3$.
 Thus the value of n is uniquely determined as 2. This result is consistent
 with the $^{51}$V-NMR analysis. As a result, a 3d$^1$ electron is
 shared within two V sites. Two values of ${\theta}_{n}$ are
 calculated to be $ \theta_n $=$( 47^\circ, 82^\circ )$, $( 19^\circ, 64^\circ )$ and $( 46^\circ,
 79^\circ )$ for the V1, V2 and V3 sites, respectively.

\section{DISCUSSION }
 The fact that a 3d$^1$ electron is shared within two V sites
 originates from relationship between hopping and Coulomb repulsions
 between neighboring sites. Formation of the CDW may be pointed out
 as another possibility. However, the insulating phase of
 $\beta$-Sr$_{0.33}$V$_2$O$_5$ is hardly explained by the nesting of
 Fermi surface; several pressure-induced insulating phases
 accompanied with superstructures were observed without the changes
 of crystal structure and carrier number. It is unlikely that all of
 the phases arise from the nesting of Fermi surface because the
 crystal structure is unchanged. There are two possibilities for a
 3d$^1$ electron shared within two V sites; one is sharing within a
 rung, and the other is sharing within a leg. In the latter model,
 the charge would be distributed to only one side of the V1 or V3
 sites according to the crystal symmetry. The latter model conflicts
 with the analysis of $^{51}$V-NMR and ZFR measurements, in which
 charge density is almost the same for the V1 and V3 sites.
 Therefore, our experimental results suggest charge sharing within a
 rung. When $\vert i>$ ($i=1$, and $2$) represents the CO state where
 a 3d$^1$ electron with up spin ($S^{z}=1/2$) is located on a site as
 shown in Fig. 14, a 3d$^1$ electron shared within a rung is
 expressed as
\begin{eqnarray}
 |a >=\frac{1}{\sqrt{2}}(|1>+e^{i\theta}|2>),
\end{eqnarray}
 where $\theta$ is an arbitrary number, but the energy depends on its
 value, as do the charge densities; an appreciable energy difference
 can be expected between the bonding state ($\theta=0$)and the
 anti-bonding state ($\theta={\pi}$). The occupation of the bonding
 state would have more charge in-between the two V sites and lead to
 a strong coupling to phonons. This would locally favor a shortening
 of the rung.

\begin{figure}
\includegraphics[width=7.8cm]{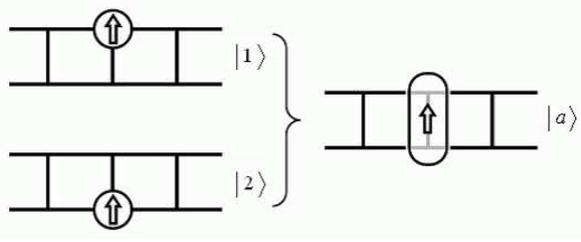}
\caption{\label{fig:epsart} spin ordering model.  $^{20}$}
\end{figure}

 The expectation values of $S_{i}^{z}$ are derived from $S_{i}^{z} \vert j>=(1/2) \delta _{ij}\vert j>$
 and $<i|j>={\delta}_{ij}$ as follows,
\begin{eqnarray}
 <a|S_1^z|a>&=&1/4 \\
 <a|S_2^z|a>&=&1/4.
\end{eqnarray}

 Note that the magnitude and direction of $<S_{i}^{z}>$ should be the
 same for the two sites. The spin directions from [1, 0, -1] are
 indicated by the sum of ${\theta}_{0}$, the angle between the first
 principle axis of EFG and [1, 0, -1], and ${\theta}_{n}$, the angle between
 the first principle axis of EFG and spin direction :
 ${\theta}_{0}+{\theta}_{n}$.  The angles are estimated as 94 and 101
 degrees for one rungs, and 145 and 156 degrees for the other rungs,
 respectively, making the difference less than about 10 degrees. The
 directions are shown in Fig. 15.
 One of the spin moment turns to the bottom plane of the VO$_5$ pyramid,
 and the other the direction of about 30 degrees from the apical direction of the pyramid.
 Similar situation holds for the V2 sites as shown in Fig. 15; The two
 directions are estimated to be 76 and 121 degrees. One may point out
 that possible directions of $<S_{i}^{z}>$ could be out of the ac
 plane. Such a situation is hardly expected if charge sharing within
 a rung is realized and the directions of $<S_{i}^{z}>$ are the same
 within a rung; Experimentally, the directions obtained independently
 for the V1 and V3 sites agree with each other only when they align
 in the ac plane.

 The spin moments with the opposite direction also satisfies Eq. (3),
 which makes the antiferromagnetic alignment possible in the AF
 phase. The existence of two kinds of spin moments for each Vi site
 is attributed to the crystallographical peculiarity in the CO phase
 : X-ray diffraction revealed the sixfold lattice modulation along
 the b axis making locally inequivalent pyramid structure on the Vi
 site. Therefore two V1-V3 rungs located on crystalographically
 inequivalent position would make two kinds of spin directions. The
 two V2s rungs are located on equivalent positions. Two sets of V2-V2
 magnetic rungs would be induced by local arrangement of the V1-V3
 magnetic rungs, although two V2-V2 rungs are located on equivalent
 positions.

 To maintain electrical neutrality, the ratio of magnetic rung and nonmagnetic rung
 should be 1 to 2. The magnetic rungs are expected to align with threefold lattice periodicity
 to avoid Coulomb repulsion. This threefold charge periodicity has also been reported for $\beta$-Na$_{0.33}$V$_2$O$_5$.
 However, the charge-distribution pattern is different;
 $\beta$-Na$_{0.33}$V$_2$O$_5$ consists of two magnetic rungs and a nonmagnetic rung.
 The reason why the charge-distribution pattern differs depending on materials is not certain at present.
 The charge distribution may be sensitive to the relationship between hopping and repulsion and as a result depends on detailed environments of the materials.

\begin{figure}
\includegraphics[width=7.0cm]{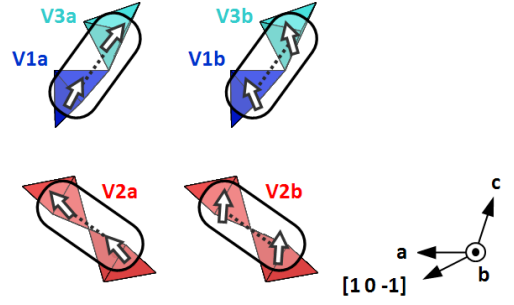}
\caption{\label{fig:epsart} The directions of $<S>$ obtained for
each Vi site independently.}
\end{figure}

\begin{figure}
\includegraphics[width=7.0cm]{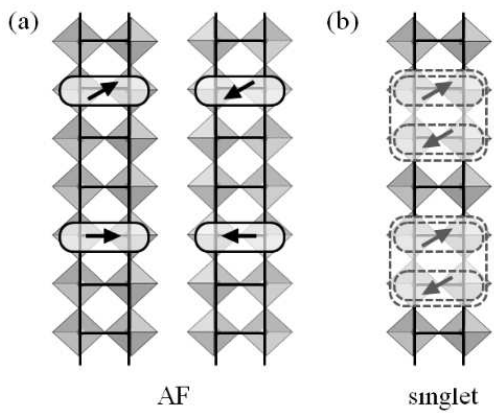}
\caption{\label{fig:epsart} Scheme of charge-sharing within a rung.
(a) Rungs with $S=1/2$ are isolated in every three lattices along
the leg due to Coulomb interaction for
$\beta$-A$^+$$_{0.33}$V$_2$O$_5$, giving a reason why the ground
state of $\beta$-Ag$_{0.33}$V$_2$O$_5$ is magnetic. (b) Two rungs
exist on the nearest positions along the leg for
$\beta$-A$^{2+}$$_{0.33}$V$_2$O$_5$, which could cause singlet
pairing. The scenario explains phenomenologically why the ground
state of $\beta$-Sr$_{0.33}$V$_2$O$_5$ is a singlet.}
\end{figure}

 The idea of charge sharing within a rung like hydrogen molecule is
 rather promising to understand why the ground state of
 $\beta$-Ag$_{0.33}$V$_2$O$_5$ is an AF ordering
 state, whereas that of $\beta$-Sr$_{0.33}$V$_2$O$_5$ is a singlet
 state. The former includes a magnetic rung with S=1/2 aligned in every three
 rungs along the leg direction (Fig. 16(a)), then coupling of
 two rungs seems difficult because such rungs would be
 located on isolated positions to avoid Coulomb interaction. On the
 other hand, the latter includes one nonmagnetic rungs in every three rungs
 along the leg direction. As a results two rungs with S=1/2 are
 located on the nearest neighboring positions (Fig. 16(b)),
 therefore singlet formation accompanied with lattice distortion
 would decrease the total energy. In real material, situation may be
 much complicated, however, essential property may be explained by
 the charge sharing.

\section{SUMMARY}
 We measured $^{51}$V NMR and ZFR on $\beta$-Ag$_{0.33}$V$_2$O$_5$ to
 investigate local properties of spin and charge for three
 inequivalent Vi sites (i=1, 2, and 3). The principal axes of the EFG
 tensor were identified for the three phases, namely the metallic phase
 at \emph{T} above and below cation ordering temperature, the
 charge-ordering phase, and antiferromagnetic odering phase. We found
 that the EFG for the V1 sites shows similar behavior to that for the
 V3 sites. The similarity was also observed as pairs of signals
 closely located in ZFR spectra. These results were explained by
 charge sharing model where a 3d$^1$ electron is shared within a rung
 in both V1-V3 and V2-V2 two-leg ladders.

\begin{acknowledgments}
 We would like to thank Y. Uwatoko, M. Hedo, N. Takeshita, T. Nakanishi, S. Yamamoto, M. Itoh, S. Fujimoto and H. Ikeda for valuable discussions.
 We also thank J. Yamaura for X-ray analysis of the crystal axes.
 This work was supported by the Japan Society for the Promotion of Science (20$\cdot$1187).
 This work was partially supported by a Grant-in-Aid (KAKENHI 17340107) from the Ministry of Education, Science and Culture, Japan.
\end{acknowledgments}


 \ \\

\end{document}